# Globally controlled artificial semiconducting molecules as quantum computers

Jerome Tribollet

*independent researcher, France*

**Quantum computers are expected to be considerably more efficient than classical computers for the execution of some specific tasks. The difficulty in the practical implementation of thoose computers is to build a microscopic quantum system that can be controlled at a larger mesoscopic scale. Here I show that vertical lines of donor atoms embedded in an appropriate Zinc Oxide semiconductor structure can constitute artificial molecules that are as many copy of the same quantum computer. In this scalable architecture, each unit of information is encoded onto the electronic spin of a donor. Contrary to most existing practical proposals, here the logical operations only require a global control of the spins by electromagnetic pulses. Ensemble measurements simplify the readout. With appropriate improvement of its growth and doping methods, Zinc Oxide could be a good semiconductor for the next generation of computers.**

Since the first theoretical work[1] on quantum computer in 1980, in which it was suggested to store a bit of information in a two level quantum system, with '0' as the ground state and '1' as the excited state , a new interdisciplinary field of research called 'quantum computation' has appeared. This interest in quantum computation has particularly increased after the discovery[2] in 1996, that quantum computers can tolerate a small amout of errors occuring during the calculation. The discovery of very efficient quantum algorithms, that could for example accelerate considerably a search through large database[3] has also stimulated lots of further studies. In the last ten years, a multitude of possible practical implementations have been suggested[4]. The most commonly used approach of quantum computation is based on individual control of the quantum bits, called 'qubits' and a list of five criteria for its practical realization is now widely accepted[5]. Quantum cellular automaton strategies reduces thoose requierements[6,7] because no individual manipulation of the nanoscopic qubits is necessary. All the qubits of an automaton are subjected simultaneously to the same evolution rule applied to the whole system using external fields. Despite the existence of several theoretical proposals, very few man-made or natural systems have been identified for the  practical implementation of such a Quantum Cellular Automaton[8,9].

 Till now, only two practical implementations have been successful in demonstrating quantum algorithms. Both of them follow the most conventionnal approach involving individual addressing of qubits. The first one is the ion trap-based quantum computer. It seems difficult to scale this computer beyond hundred qubits. The second is the nuclear magnetic resonance-based quantum computer[4,10,11]. This latter quantum computer has the enormous advantage that the readout of the results of the quantum computation is a simple average magnetization measurement over a macroscopic number of clones of the same quantum computer (molecules) that run in parallel. However, for intrinsic reasons related to nuclear magnetic resonance[12], this proposal is not scalable. To solve this problem, Kane has proposed a solid state planar version of this computer[13], further modified by Vrijen[14]. In this last version, the qubits are encoded onto the electronic spins of phosphorous donors embedded in a silicon matrix and manipulated by individual gates. However, the planar array of closely spaced donors and gates required to build thoose quantum computers remains extremly difficult to obtain in practice.

Taking advantage of the fact that hydrogen atoms are shallow donors in Zinc Oxide[15,16], but also of well developed semiconductor growth methods, i describe how it would be possible to build an ensemble of parallel vertical lines of hydrogen donors in this semiconductor. Thoose donor lines constitutes the electronic spin chains used for quantum computation. Virtual excitations of excitons by lasers in the structure creates bandgap-dependant effective magnetic fields[18,19] but also bandgap-dependant effective spin-spin couplings[18,20] between nearest neighbor donor atoms along the molecules. It is also shown that at low residual donor concentration, very long electronic spin coherence time are expected for hydrogen donor atoms in Zinc Oxide. Initialization and readout of the computer can be viewed as a time resolved pump-probe Faraday rotation experiment[21,22,23]. Manipulation of the molecules that constitutes as many clones of the same quantum cellular automaton, is performed between initialization and readout, using synchronized microwave and optical pulses.

**Opportunity for an improved doping technology in ZnO**

Using available molecular beam epitaxy growth methods (MBE), it is now possible to build semiconductor heterostructures whoose composition is controlled atomic layer by atomic layer. The resulting bandgap profile engineering is now available for nearly all types of semiconductors, like silicon[14] (Si), gallium arsenide (GaAs) and also for semiconductors of the group II-VI like Zinc Oxide[24] (ZnO). Silicon has been identified has the best semiconductor for the implementation of a spin-based solid state quantum computer because it has the longest electronic spin coherence time and because silicon technology is well developped for classical computers[13,25]. However, it seems probable that ion implantation methods commonly used for silicon doping will never reach the atomic level of precision required to build silicon based quantum computers[14]. Recently, a new method to introduce phosphorous donors in silicon has been proposed that takes advantage of well developped hydrogen atoms manipulation methods using the tip of a scanning tunneling microscope[26,27] (STM). However, this method remains quite complicated because it implies the controlled adsorption of $PH_3$ molecules to STM-patterned sites of H-terminated silicon surfaces and a subsequent annealing step. Furthermore, this method suffers from a very low fabrication rate. To increase the fabrication rate, a possible solution would be to use arrays of STM tips operating in parallel. However till now, arrays of STM tips contains only tens or hundreds of tips, spaced typically by one hundred microns[28].

The first idea that i propose in order to develop an easier doping technology consist in using Zinc Oxyde (ZnO) as the host semiconductor for the quantum computer instead of silicon. One of the reasons is that it was recently proposed theoretically[15] and verified experimentally[16], that hydrogen (H) is a relevant shallow donor atom for ZnO. This opens a unique opportunity to use well developed manipulation methods of hydrogen atoms by STM to directly control the doping on clean ZnO surfaces. The second idea that i propose aims to improve the fabrication rate of STM assisted doping methods. It consist in building an ultra dense array of homogeneous STM tips that could be used in parallel. For that, I suggest to take advantage of a new nanofabrication method assisted by lasers[29,30]. Appropriate standing waves of lasers placed above a given substrate can act on neutral atoms as an array of 'atom lenses' or a 'light mask', that focuses the falling atoms into small areas to create the desired structure. The typical spacing between thoose 'atom lenses' is half the wavelength of the laser used to manipulate a given atom, that is typically several hundreds nanometers. Two dimensionnal arrays of nanoscopic chromium dots have been recently obtained[31]. This array of chromium dots is roughly a square of 200 µm x 200 µm , with a typical spacing between two adjacent dots of 200 nm, all the heigths of the dots being equal with a precision of one nanometer. I propose to consider this dense array of dots as an array of one million STM tips that could act in parallel. As we only want to desorb ('vertical processes'[26]) and then to redeposit hydrogen atoms in a massively parallel manner and not to image the surface to be doped, a single 'z' feedback loop common to all the tips of the array is required to control the vertical distance between the tips and the surface ('z' direction). A single pair of 'xy' controllers is also required for the entire tip array in order to control its lateral displacement above the surface of the substrate. The resulting new doping method for ZnO can be summarized in the following manner. Typically two chambers with ultra high vacuum will be used. One is used for the growth by molecular beam epitaxy of the ZnO structure. The other must contains the tip array and an H-terminated surface of silicon that will serve as a 'solid source' of hydrogen atoms for doping. At each doping step of the ZnO structure, the MBE growth is stopped in the first chamber. In the second chamber, hydrogen atoms are desorbed in parallel from the hydrogeneted surface by the tips and then transfered with the whole array in the first chamber, where hydrogen atoms are finally deposited on the ZnO surface. By this way, an array of one million donor atoms on ZnO is created. It can be embedded in the structure by further MBE growth steps, after the tip array has been retired from the first chamber. By superposing vertically many donor arrays of this type, we create one million of identical donor chains that can be identified as artificial molecules usefull for the practical implementation of a quantum cellular automaton.

**Band gap engineering and manipulation of the artificial molecules**

The quantum cellular automaton that i propose to build is the one described few years ago by Lloyd[6]. It can be viewed as a periodic linear artificial molecule of the type ABCABC...ABCABC. The letters correspond to different types of segments of the molecule (called units or cells or quantum bits) that differ essentially by their spectral properties. Here, the cells are distinguished both in the optical range and in the microwave range by an appropriate bandgap profile and doping engineering as indicated on Fig. 1. Each cell here is a single hydrogen donor contained in the middle of a quantum well with a given bandgap profile. The electronic spin ½ of a given donor is a two states quantum system that endoce a quantum bit (states noted '0' and '1' are the 'down' and 'up' spin states respectively). Each quantum bit of this molecule can be in a superposition of the up and down spin states as required for quantum computation. The basic ideas behind the Zn$_{1-x}$Mg$_x$O / Zn$_{1-y}$Mg$_y$O/... /Zn$_{1-w}$Mg$_w$O heterostructure[24] of Fig. 1, containing undoped quantum wells and doped in the well quantum wells, are the following. First, it must be noted that, if the hydrogen donor atoms would be deposited in a bulk ZnO structure, their electronic spins would all have nearly the same g-factor and then the same Zeeman splitting in the presence of an external magnetic field B$_0$ applied along the growth axis of the structure (the c-axis of wurtzite ZnO, also here, the 'z' axis of quantification of the spins). A simple solution to this spectral problem would consist to embbed donors in quantum wells that have very different bandgap profile and to take advantage of the bandgap dependance of the electronic g-factor[14] to distinguish the different types of cells of the molecule in the microwave range under an applied magnetic field B$_0$. However, to obtain very different Zeeman splittings at reasonable magnetic field B$_0$ in the ZnMgO heterostructure, the depths of the different types of quantum wells have to be very different because the electronic g-factor is only weakly dependent on it. If on one side, this solution provide cells distinguished in the microwave range as required to implement the Lloyd quantum computer with electronic spin-based qubits, on the other side, nearest neighbor coupling between two successive donor electronic spins in the molecule becomes impossible. The reason is that we have only two choices to implement the coupling in this case. The first choice consist in using the dipolar spin-spin coupling and the second to use the direct exchange coupling (in the growth direction). However, in order to obtain a sufficiently strong spin-spin coupling necessary to implement the Lloyd quantum computer based on spectral selectivity, two successive doped quantum wells would have to be separated by a very small distance. Then, all the electrons of the donors contained in the wells that have small well depths would tunnel irreversibly towards the doped quantum well that have a larger well depth. So, we conclude that g-factor engineering is not compatible with spin-spin coupling in a vertical heterostructure.

The solution that i propose consists in introducing an appropriate undoped quantum well ('R' as Relay) between each pair of successive doped quantum wells, as indicated on Fig. 1, in order to increase the inter-donor distance and then to suppress tunneling. In applying two simultaneous appropriate laser pulses to the whole heterostructure, we obtain simultaneously different spin splittings ($\Delta_i = \Delta_A, \Delta_B$, or $\Delta_C$) for the different types of donors (cells) and different exchange mediated spin-spin couplings ($J_{ij} = J_{AB}$, $J_{BC}, J_{CA}$) between two successive types of donors in the molecule. The optical pulses propagate in the z direction (growth direction). The effective magnetic fields and spin-spin couplings exist only during a time equal to the common duration of the two optical pulses, as a result of the virtual creation of excitons in the different types of quantum wells, doped and undoped, in the heterostructure. One of the two optical pulses is linearly polarised (laser energy $\omega_L$ and Rabi energy $\Omega_L$, with, $h/(2\pi) =1$) and produces the dominant exchange coupling between each pair of successive donors vertically aligned in the molecule[18,20]. The other optical pulse is circularly polarised (laser energy $\omega_C$ and Rabi energy $\Omega_C$) and creates the different spin splittings[18,19]. The spin splittings (equivalent to effective magnetic fields) and spin-spin couplings are strongly dependent functions of the laser frequencies and bandgap profile. This is the key point to obtain the spectral selectivity in the microwave range. The simplified interband optical absorption spectrum corresponding to the qualitative bandgap profile of Fig. 1 is indicated on Fig. 2. On this optical absorption spectrum, it appears both free exciton transitions (noted X) and 'donor-bound exciton' transitions[32] (noted D0X) associated to the different types of quantum wells in the structure. The undoped quantum well 'R' is characterized by its exciton 1S transition $X_R$ (with an optical pulsation $\omega_R$). Each doped quantum well (i=A,B, or C) is characterized by an excitonic 1S transitions ($X_i = X_A$, $X_B$, or $X_C$), but also by a fundamental 'donor-bound exciton' transition ($D0X_i = D0X_A$, $D0X_B$, or $D0X_C$). The dominant spin-spin coupling created by the linearly polarised optical pulse depends on the detuning $\delta_R = \omega_R - \omega_L$. The spin splitting of a donor in a given quantum well created by the circularly polarised optical pulse depends on the detuning $\delta_i = \omega_{D0Xi} - \omega_C$. At large detunings, the spin splittings[19] vary as, $\Delta_i = (\Omega_C)^2 /\delta_i$, and the spin-spin exchange couplings of Heisenberg type (ORKKY interactions[20]) vary as,

$J_{ij} = J_{iR}.J_{Rj} .(\Omega_L)^2 / (\delta_R)^3$, where $J_{iR}$ and $J_{Rj}$ are the vertical exchange couplings between the exciton in the undoped quantum well of type R and the donors in the adjacent doped quantum wells, i and j respectively. As we want to manipulate the quantum bits encoded onto the electronic spins of donor atoms, we need to use a pulse EPR spectrometer[8,33] to apply microwave pulses on the ZnMgO structure. A priori, the microwave resonator works at a fixed frequency. We choose a central microwave frequency at

$\nu_{CAV}$ = 23 GHz (K-Band, [18GHz, 26 GHz]), corresponding to a microwave photon energy $\Delta_{CAV}$ = 95 µeV. The typical set of parameters of the ZnMgO heterostructure is, $\delta_A$ = 12 meV, $\delta_B$ = 10 meV , $\delta_C$ = 8 meV (this defines the energy of the circularly polarised laser). The energy offset, $\varepsilon$ = 2 meV (see Fig. 1 and Fig. 2), between two successive D0X transitions can result either from different compositions or from different widths of the wells of the doped quantum wells. If we apply a circularly polarised optical pulse with a Rabi energy of $\Omega_C$ = 1.07 meV, we obtain $\Delta_A$ = 95 µeV, $\Delta_B$ = 114 µeV, $\Delta_C$ = 143 µeV. In this case, $\Delta_A = \Delta_{CAV}$ = 95 µeV and then only the donors of type A of the molecule can be manipulated by the microwave pulse. By adjusting the value of $\Omega_C$ we can control the type of donor electronic spin in the molecule that we want to manipulate with a microwave pulse (A, B, or C). The typical excitation profile of a selective 'π' pulse[6] at K-band, with a typical duration of 10 ns, has a width in energy on the order of $\delta_{EX} = 0.06$ μeV (for a Fourier-transform gaussian pulse for exemple). For $|J_{ij} - J_{jk}| > \delta_{EX}$ and $\Omega_C$ close to 1.07 meV, it is possible to perform the selective manipulation of the spin state of all donors of type A, conditionned to the spin states of their two nearest neighbour donors, one B and one C, as a result of the spin-spin couplings. For example, and following the notation of Lloyd[6], if we suppose that all the donors of the molecule ABCABC...ABCABC are initially either in the spin up state '1' or in the spin down state '0' (case were the molecule is used as a classical cellular automaton) and if a microwave 'π' pulse called $\omega^A_{01}$ is applied to the molecule, then all the donors of type A of the molecule that have a C= 0 donor on the left and a B= 1 donor on the right will switch their spin state from 0 to 1 or from 1 to 0, depending on their initial spin state. This switch, or spin flip, either results from the coherent absorption or emission of a microwave photon at the resonant energy $\Delta_{CAV} = \Delta^A_{01} = \Delta_A + J_{BA} - J_{AC}$. In this case, $\Omega_C$ is different from 1.07 meV to put the transition $\Delta^A_{01}$ in resonance with the fixed photon energy of the microwave pulse in the cavity $\Delta_{CAV}$. In principle, the effective exchange interaction between donors can be estimated by $J_{ij} = J_{iR}.J_{Rj} .(\Omega_L)^2 / (\delta_R)^3$, as we already said above. However, it was recently shown[18] that due to the possible binding of an exciton on a neutral donor, this effective exchange can have a complex resonant behavior as a function of the laser detuning from the appropriate excitonic transition. More precisely, if the laser energy $\omega_L$ is close to the resonance of a 'donor bound exciton' transition, we expect a huge enhancement of this effective exchange coupling with a maximal value on the order of the Rydberg energy in the semiconductor considered. In ZnO, the bulk exciton binding energy $E_b(X)$ is equal to 60 meV. As the theory of Piermarocchi & al. is already semi-phenomenological and needs some experimental data to be quantitative, we don't go further in the estimation of $J_{ij}$. We just stress that if $\omega_L$ is slightly below the 'donor bound exciton' transitions of the ZnMgO heterostructure (few meV), we expect that $J_{ij}$ be on the order of several µeV and that $|J_{ij} - J_{jk}| > \delta_{EX}$ as required. In this case, the spin-spin couplings becomes effectively of Ising type, because due to the other circularly polarised laser, we have $|\Delta_j - \Delta_i| \gg J_{ij}$ (for i= A, B or C). It is interesting to note that the virtual exciton in the undoped quantum well plays a role analog to the 'J' gate in the Kane quantum computer[13] in reducing the effective barrier between donors (see Fig. 1). All the requierements for the implementation of the conditional quantum dynamic of the molecule seems then achievable with a carefull adjustement of the laser energies. When no virtual excitons are present in the R-type undoped quantum wells, two adjacent donors in the growth direction are too much confined and spaced in order that tunneling be possible. Note also that the two optical pulses and the microwave pulse must have the same length (typically 10 ns) and be simultaneously applied to the molecule to perform the conditionnal evolution. In the most general case, the molecule can be used as a quantum cellular automaton, simply by adjusting the duration and the phase of the microwave pulses applied in order to perform any required unitary evolution (single spin rotation of a given type of donor electronic spin conditionned to the spin states of the neighbours)[33]. It must also be stressed that ZnO is particularly well suited for the practical implementation of this quantum cellular automaton using 'optical gating' because the exciton oscillator strength in ZnO is very large, at least two order of magnitude larger than in GaAs[34]. This allows to obtain laser Rabi energies of few meV during 10 ns as requiered.

### Decoherence in ZnO

Before to study the spin decoherence processes in ZnO, we have to consider the possible inhomogeneous broadening of the Electron Spin Resonance lines (ESR) in the ZnMgO structure proposed. This inhomogeneous broadening results mainly from the linewidth of the 'donor bound exciton' transitions in the optical range, noted $\Gamma^{D0X}$, because the optically induced spin splittings depend on the detuning between the 'donor bound exciton' transition and the photon energy of the circularly polarised laser. Thus, the inhomogeneous broadening of an ESR line can be estimated by:

$\Gamma^{ESR}_{inh, i} = \Gamma^{D0Xi} . (\Delta_i / \delta_i)$. To estimate this inhomogeneous broadening we need an estimation of $\Gamma^{D0X}$. In a sufficiently clean structure, $\Gamma^{D0X}$ is limited by the homogeneous optical linewidth of the 'donor bound exciton' transitions. Then, $\Gamma^{D0X}$ is inversely proportional to the recombination time of an exciton[35] bound on a neutral donor, $T_{REC}$. In ZnO, $T_{REC} = 60$ ps. It corresponds to $\Gamma^{D0X} = 10$ µeV. With the typical values, $\delta_A = 12$ meV and $\Delta_A = 95$ µeV, we find an inhomogeneous broadening of the ESR lines of the order $\Gamma^{ESR}_{inh} = 0.08$ µeV, that is comparable to $\delta_{EX} = 0.06$ µeV as requiered for a full excitation of a given ESR line by the microwave pulse. To further reduce this inhomogeneous broadening, a solution consist to decrease the photon energy of the circularly polarised laser (to increase $\delta_i$) and to decrease its Rabi energy (to reduce $\Delta_i$), eventually adding an external magnetic field $B_0$ in the growth direction to adjust the resonance condition with the microwave cavity. An other solution consist to use other type of selective microwave pulses that have a larger excitation bandwidth $\delta_{EX}$.

In a bulk ZnO crystal with a residual donor concentration of $10^{17}$ cm$^{-3}$, Baranov[16] and al. have measured an halfwidth $\Delta B_{1/2} = 0.2$ mT for the electronic spin resonance line associated to hydrogen donors in ZnO, corresponding to $T_{2,exp} = h /(4\pi\, g_{av}\, \mu_B\, \Delta B_{1/2} ) = 13$ ns. From the ionisation energy of the hydrogen donor measured in this sample, $R_0 = 35$ meV, we deduce a critical donor concentration for the metal-insulator transition[17] of $7.10^{17}$ cm$^{-3}$. Then, we expect that both insulating and metallic clusters of donors exist in this sample. In both phases, the decoherence of the electronic spin of a donor can be explained by a spin-orbit hamiltonian called the 'k-linear term' and written in the form of equation (1):

$$< H_{SO} = C_0 (k_x\, \sigma_y - k_y\, \sigma_x) \qquad (1)$$

where $C_0 = -64$ µeV.nm, is the 'spin-orbit/crystal field' coupling constant[36,37] in ZnO and where $\sigma_x$ and $\sigma_y$ are the x and y Pauli matrix. In the metallic phase, the 'k-linear term' acts on the spin as an effective magnetic field contained in the xy plane orthogonal to the c-axis of the wurtzite ZnO crystal. The direction of this effective magnetic field change at each collision because the wave vector 'k' of the delocalized electron change at each collision and then induces the decoherence of the electronic spin. Taking advantage of the formal analogy between the 'k-linear term' and the 'Rashba term' generally met in asymmetric semiconductor quantum wells of group III-V or group IV semiconductors[38,39], we can predict for hydrogen donors in ZnO, a g-factor anisotropy $\Delta g$ and a decoherence time $T_{2,metal}$ of the electronic spin given respectively by equation (2) and equation (3):

$$\Delta g = 0.5\ (C_0\, k_F)^2 / (g^*\, \mu_B\, B_0)^2 \qquad (2)$$

$$T_{2,metal} = 0.5 \, (1/\tau_C) \, h^2 / (2\pi \, C_0 \, k_F)^2 \qquad (3)$$

From the measured mobility $\mu = 2000 \text{ cm}^2 \text{ V}^{-1} \text{ s}^{-1}$, we deduce a collision time $\tau_c = \mu.(m^*/e)$ of 340 fs in this sample. From the measured g-factor anisotropy $\Delta g = 0.0017$ ($g_{av} = 1.956$, and $B_0 = 3.467$ T), we deduce a local Fermi wave vector $k_F = 0.359 \text{ nm}^{-1}$, from which we deduce a decoherence time $T_{2,metal} = 1$ ns. Still using the analogy with the 'Rashba term', we predict a decoherence time of the donor electronic spin in the insulating phase[40,41] given by equation (4):

$$T_{2,ins} = \beta(T) \, h / (n_0 \, a_0^3 \, R_0 \, \gamma^2) \qquad (4)$$

where $\beta = 0.05$ at T= 4.2 K, $n_0$ is the donor concentration, and $\gamma$ is the average rotation angle estimated by[42], $\gamma = C_0 / (R_0.a_0) = 8.5 \, 10^{-4}$. For a donor concentration $n_0 = 10^{17} \text{ cm}^{-3}$ we obtain a decoherence time $T_{2,ins} = 20$ μs. We verify that the measured[16] decoherence time of 13 ns is comprised between the two estimated decoherence times. The practical implementation of the proposed quantum computer requires T= 4.2 K and $n_0 = 10^{16} \text{ cm}^{-3}$ to obtain $T_{2,ins} = 200$ μs. In this range of concentration (clearly insulator) and for initially spin polarised nuclear spins of the host ZnO (by optical pumping), the hyperfine interactions between the hydrogen donor and the surrounding nuclei limit the decoherence time[42] to $T_{2,limit} = N.(h/A_{Zn/H}) = 90$ μs (with[16] N=50 nuclei, $A_{H/H} = h \, \nu_{H/H}$, and $\nu_{H/H} = 1.4$ MHz). For a gating time $T_p = 9$ ns, we then calculate a ratio $T_{2,opt} / T_p = 10^4$ that is sufficient to implement fault tolerant quantum computation[5]. Note that a small static magnetic field of around hundred gauss will have to be applied to the system to avoid other decoherence mechanisms induced by nuclei[43], efficient at zero external magnetic field. Spin-orbit processes assisted by phonons[44] are irrelevant in ZnO at T= 4.2 K.

**Initialisation, readout and error correction .**

Recently, time-resolved Faraday rotation experiments[23] have been performed on ZnO crystals. In this kind of pump-probe experiment[22], a first optical pulse, circularly polarised (the pump), is absorbed by the semiconductor and results in the creation of spin-polarised real carriers. Then, after a controlled time-delay, a second optical pulse, linearly polarised (the probe), is transmitted through the sample and then its polarisation is analysed using a polariser or a more sensitive device. The rotation of the direction of polarisation of the probe pulse is proportional to the component of the sample magnetization ($M_z$) along the direction of propagation of the probe pulse (along the c-axis of the ZnO crystal, also called here the z axis). I suggest to consider this experiment as a 'writing-readout' experiment.

Initialisation of all the spins of the ensemble of artificial molecules contained in the ZnMgO heterostructure can be obtained by optical pumping using a circularly polarised light pulse resonant with the continuum of excitation of the molecules. The idea dehind this strategy is that it is expected that free photocreated holes relax their spins more rapidly than free photocreated electrons before recombining onto the donors. A $\sigma^+$ polarised optical pulse for example will initialise all the molecules (chains of donors) to the simple state where all the donor electronic spins are down. In practice, this initialisation will require a time on the order of the recombination time of the photocreated carriers, typically hundred picoseconds. It is possible to apply this optical pumping step during a longer time before starting the quantum computation, to be quite sure that all the artificial molecules are appropriately spin-polarised.

Note that a trivial solution to initialize the molecules to the ground state would be to cool them at low temperature under a strong external magnetic field, but this solution requires either a strong external magnetic field difficult to switch rapidly (necessary if we want to work in the K-band) or very low temperatures (100mK) difficult to combine with an optical acces to the sample. Then optical pumping is the simplest and fastest solution for the initialization of this quantum computer. The appropriate sequences of global evolution rules necessary to load and unload information onto this molecule have been described by Lloyd[6]. At the end of the quantum computation, we want to read the states of all the encoded qubits in the molecule. For that, we transfer the state of the encoded qubit that we want to read towards the end of the molecule using an appropriate sequence of pulses. In order to simplify the readout process, a last donor of type D was added at the end of the molecule that becomes ABCABC...ABCABCD. This last type of donor is spectrally distinguished from the others, both in the optical range and in the microwave range, following the same strategies as for the three other types of donors (see Fig. 1 et Fig. 2). The advantage of this D-type donor at the end of the molecule is that its 'donor-bound exciton' transition is the lowest in energy among all the 'donor-bound exciton' transition of the molecule. Then, a Faraday rotation measurement of $M_{z,D}$ with a linearly polarised optical pulse quasi-resonant with the 'donor-bound exciton' transition associated to the donors of type 'D' will provide a measure of the state of the qubit that was previously transfered on the last D cell. Note that before performing the Faraday rotation measurement, it is possible to rotate the quantum state contained on D-type donors with appropriate microwave pulses. By this way, and in repeating many times the same computation, the full quantum state tomography of each encoded qubit can be performed, in analogy to the bulk NMR quantum computer[4,10]. It is important to note that ensemble measurements over a great ensemble of clones of the same molecule greatly simplify the readout process and overcome the challenge of single spin measurement[9]. Note that if probeheads are also available in the pulse EPR spectrometer, an ensemble measurement in the microwave range over the million of molecules contained in the sample is possible, in exact analogy to the bulk NMR quantum computer.

**Outlook.**

Taking advantage of the recent discovery that hydrogen atoms acts as shallow donors in the wide gap Zinc Oxyde semiconductor[15,16], I have proposed a scalable quantum computer architecture based on a ZnMgO heterostructure. This quantum computer is globally controlled. It means that initialization, quantum information processing and readout are all performed using externally applied electromagnetic fields, here in the optical and microwave range. This removes many technological problems generally met in other solid state quantum computer proposals like individual manipulation and readout of a single spin at nanoscopic scale.

The practical implementation of this architecture is based on the possibility to introduce arrays of donors inside ZnO quantum wells. This type of requierement is found in many other solid state quantum computer proposals[13,14] and remains difficult to reach with available semiconductor doping technologies. A possible solution to this problem is to use very well developped methods for the manipulation of hydrogen atoms with the tip of a scanning tunneling miscroscope[26] (STM). In a ZnMgO structure where hydrogen is directly a shallow donor, the 'doping challenge' is then reduced to the creation of a dense array of STM tips that could manipulates many hydrogen atoms in parallel. Recently improved nanofabrication methods assisted by laser could be an efficient solution to build the required STM tips array. The conventional 'delta-doping' method already available could be used for first test-experiments like the calibration of the spin splittings and exchange couplings between donors as a function of the laser detuning and Rabi energy, eventually in other II-VI or III-V semiconductor quantum wells, delta-doped inside the well.

In order to obtain a long electronic spin coherence time for shallow donors in a ZnMgO heterostructure grown by molecular beam epitaxy, some efforts will have to be made during the growth to avoid unwanted n-type doping of ZnMgO by hydrogen and halogen atoms. A residual donor concentration equal or less than $10^{16}$ cm$^{-3}$ is predicted to be sufficient for fault tolerant quantum computation with ZnO. For first test-experiments, p-type modulation doping of the barriers could be used to effectively reduce the n-type doping in the quantum wells.

As a renewed interest is observed since few years for ZnO (p-type doped[45] ZnO and Mn-doped[46] ZnO), it is expected that the efforts made to improve the quality of epitaxial layers of ZnO, in the perspective of 'classical' spintronics applications[47], could also contribute in the future, to the development of 'quantum computers' based on ZnO.

Correspondence should be addressed to jerome Tribollet. (e-mail: jerome.tribollet@tiscali.fr)

Figure1- Qualitative conduction band profile of the periodic ZnMgO heterostructure. The structure contains doped 'in the well' quantum wells (A, B, C, D) and undoped quantum wells (R). The virtual excitation by laser of excitons ($X_v$) in the undoped quantum wells mediates an effective exchange interaction along the growth direction z between two successive confined donors ($D_0$). $\varepsilon$ and $\mu$ are energy offsets typically on the order of few meV. $\varepsilon$, $\mu$ and the inter-quantum well distance (few donor Bohr radius) control the maximal strength of this exchange interaction (adjusted by the laser detuning). Unwanted tunneling from doped towards undoped quantum wells is avoided due to the large hydrogen donor binding energy in bulk ZnO, $E_b(D_0)$= 35 meV and due to $V_A$, the potential depth of type-A quantum wells. When $V_A$ increases ($V_B=V_A+ \varepsilon$, $V_C=V_B+ \varepsilon$, $V_D=V_C + \varepsilon$), all the binding energies increase (donor: $E_b(D_0)$, exciton: $E_b(X)$).

Figure2- Optical absorption spectrum of the ZnMgO heterostructure. This spectrum contains mainly nine lines (low energy side). Four 'donor-bound exciton' transitions ($D_0X$) associated to the four types of doped quantum wells of the structure (a, b, c, d, separated by an energy offset $\varepsilon$) and five excitonic transitions (X). $E_b(Do/X)$ is the binding energy of an exciton on a donor in ZnO (at least 10 meV). The four excitonic transitions situated at an energy $E_b(Do/X)$ above the 'donor-bound exciton' transitions are associated to free excitons in the doped quantum wells (a, b, c, d). A fifth excitonic transition, with a transition energy slightly below the excitonic transition of type-D quantum wells (energy offset $\mu$ below $X_D$) is also observed (noted $X_R$). This line correspond to free exciton absorption by the undoped quantum wells. Two non resonant lasers, one linearly polarised (energy $\omega_L$) and an other circularly polarised (energy $\omega_C$) establish the spin splittings of donors and their spin-spin couplings without creating any real carrier. The intensities of the splittings and couplings depend on the different energy detuning $\delta_R$, $\delta_A$, $\delta_B$, $\delta_C$, $\delta_D$, as described in the text.

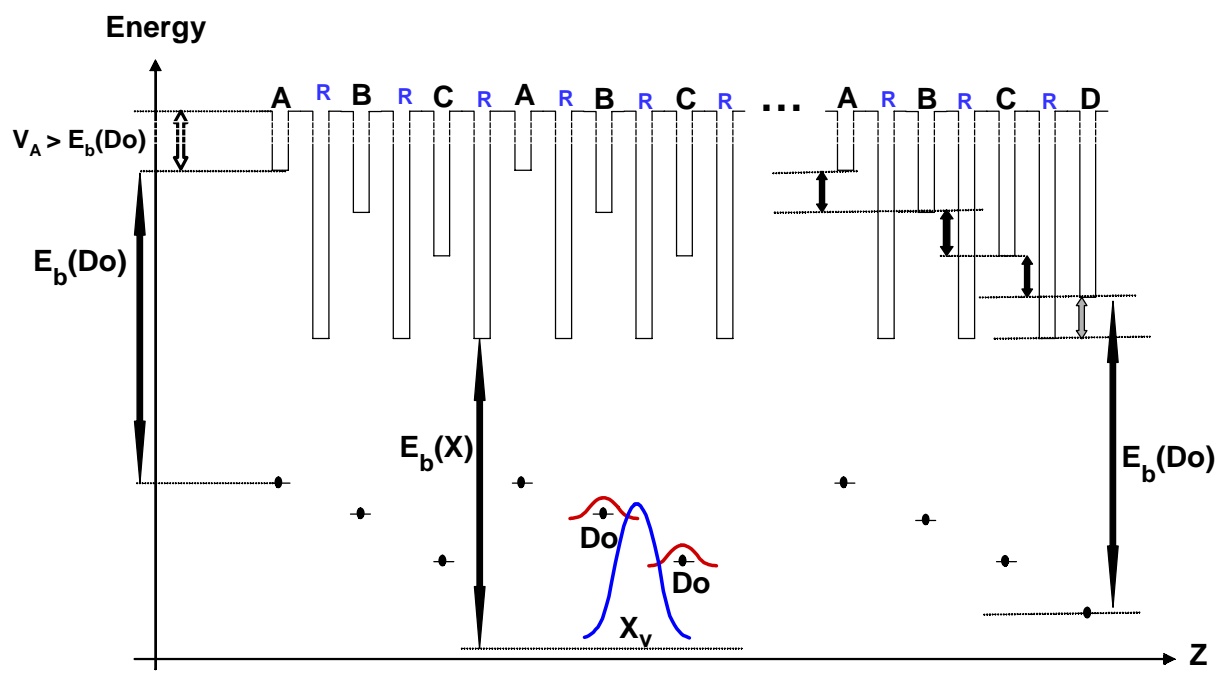

Tribollet_fig1

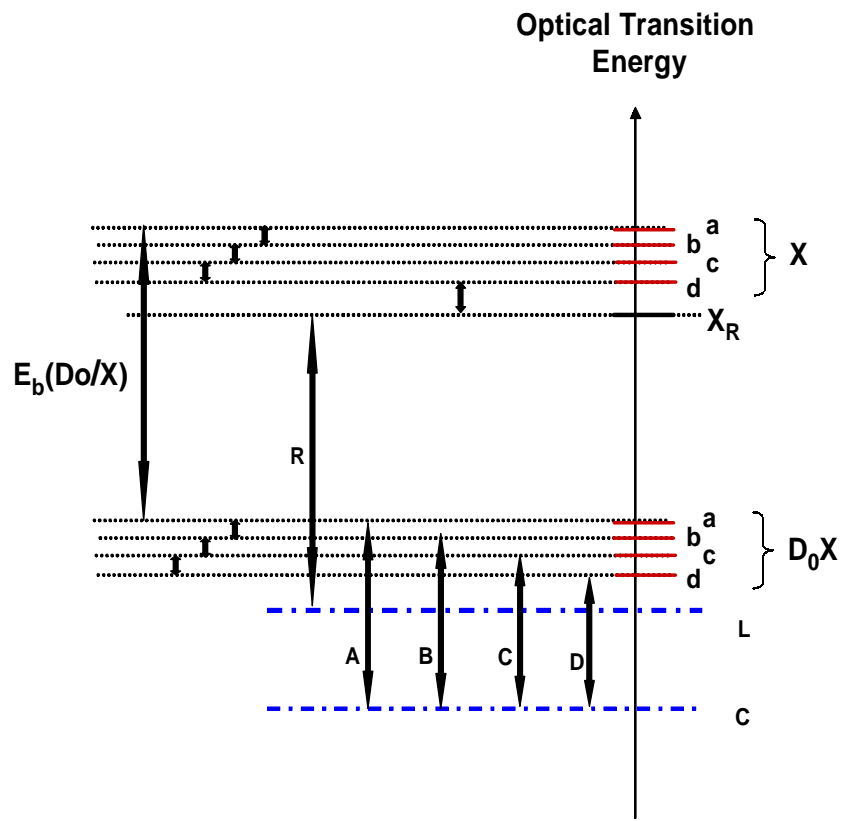

Tribollet_fig2